# Design and Description of the CMS Magnetic System Model

Vyacheslav Klyukhin [1,2]

1. Skobeltsyn Institute of Nuclear Physics, Lomonosov Moscow State University, RU-119992 Moscow, Russia; Vyacheslav.Klyukhin@cern.ch; Tel.: +41227676561
2. CERN, CH-1211 Geneva 23, Switzerland

**Abstract:** This review describes the composition of the Compact Muon Solenoid (CMS) detector and the methodology for modelling the heterogeneous CMS magnetic system, starting with the formulation of the magnetostatics problem for modelling the magnetic flux of the CMS superconducting solenoid enclosed in a steel flux-return yoke. The review includes a section on the magnetization curves of various types of steel used in the CMS magnet yoke. The evolution of the magnetic system model over 20 years is presented in the discussion section and is well illustrated by the CMS model layouts and the magnetic flux distribution.

## 1. Introduction

The discovery of the Higgs boson [1–3] with a mass of 125 GeV/$c^2$ became possible due to an observation of the signals in two decay channels of this particle produced in proton–proton interactions. The first channel with a probability of 0.2% [4], is the decay channel of the Higgs boson (*H*) [5–7] into two gamma rays: $H \to \gamma\gamma$. This signal is hardly distinguished from the prevailing background of the numerous electromagnetic decays of hadrons. The second decay channel, called "gold", is a 2.7% [4] probability decay of the Higgs boson into two *Z* bosons, one of which is virtual, with their subsequent decays into two oppositely charged leptons each: $H \to ZZ^* \to 4l$. In this case, by leptons *l* we mean electrons (positrons) *e* and muons *μ*. In the Higgs boson invariant mass reconstruction, four-momenta of leptons are used. This requires not only measuring the three-dimensional momenta of leptons, but also a reliable identification of these particles.

In modern detectors at circular accelerators, in particular at the Large Hadron Collider (LHC) [8], the precision tracking detectors placed in a magnetic field are used to measure momenta of the charged secondary particles generated in the interactions of the accelerated primary particle beams. The magnetic field gives to the particle trajectories in the tracking detector a curvature [9], which depends on the magnetic flux density ***B***. A higher magnetic flux density provides a larger curvature and, as a result, a more accurate measurement of the momentum of a charged particle. Electrons and muons are identified using other systems of the experimental setup—an electromagnetic calorimeter and a muon spectrometer [10,11]. As a rule, the superconducting solenoids with a central magnetic flux density of 2–4 T [12–19] are used to create a magnetic field in modern experimental setups at the circular colliders.

During the preparation of proposals for multi-purpose experiments [20–24] on the search for new particle phenomena at the circular colliders with the centre-of-mass energies of colliding beams of 6–14 TeV, the processes of the Higgs boson production with a mass of up to 1 TeV/$c^2$ with its decay into leptons in the final state were studied [25–28]. For a given mass of the Higgs boson, the statistical significance of the signal from its production remains constant in the region of transverse lepton momenta of 50–100 GeV/$c$.

Therefore, the measurement of the lepton momentum in this region should occur in a tracking detector with a high accuracy. It was shown that the experimental resolution across the transverse momentum of a charged particle is not only directly related to the magnetic flux density in the volume of the tracking detector, but also is determined by the degradation of the double integrals of the magnetic field along the particle paths in the extreme regions of the cylindrical volume of the tracking detector [29,30].

Modern magnetic systems of multi-purpose detectors at the circular colliders are mostly heterogeneous [18,19,31,32], i.e., the magnetic flux they create penetrates both non-magnetic and ferromagnetic materials of the experimental setup. The steel yoke of the magnet is used, as a rule, as magnetized layers wrapping muons, which makes it possible to identify them and measure their momenta in a muon spectrometer. In a solenoidal magnet, the large volume of steel yoke around a coil and the inhomogeneity of the magnetic flux density in the yoke make direct measurements of the magnetic field inside the yoke blocks difficult. Existing methods for measuring the magnetic field using Hall sensors or nuclear magnetic resonance (NMR) sensors are successfully applied within the volume of the solenoidal magnet and provide a high measurement accuracy. To use such sensors in measurement of the magnetic flux density $B$ inside the blocks of the magnet steel yoke, thin air gaps cutting the blocks in planes perpendicular to the magnetic field lines are needed. This greatly complicates the yoke as a supporting structure. An alternative option is to use a magnet current ramp down from the operating value to zero and to integrate over time the electrical signals induced by changes in the magnetic flux in sections of special flux loops installed around the yoke blocks. In the latter case, as a result of integration, the initial average density of the magnetic flux in the cross section of the flux coil can be reconstructed.

Both options allow only discrete measurements of the magnetic flux distribution in the magnet steel yoke. These are insufficient to measure muon momenta in a muon spectrometer. A computing modelling of the magnetic system to obtain the distribution of the magnetic flux throughout the entire experimental setup achieves this goal.

## 2. Materials and Methods

### 2.1. The CMS Detector Description

In the Compact Muon Solenoid (CMS) detector [11] at the LHC [8], the magnetic field is provided by a wide-aperture superconducting thin solenoid [33] with a diameter of 6 m and a length of 12.5 m, where a central magnetic flux density $|B_0|$ of 3.8 T is created by an operational direct current of 18.164 kA [34–36]. The CMS multi-purpose detector, schematically shown in Figure 1, includes a silicon pixel tracking detector [37], a silicon strip tracking detector [38], a solid crystal electromagnetic calorimeter [39] to register $e$ and $\gamma$ particles, and a hadron calorimeter of total absorption [40] both located inside the superconducting solenoid, as well as a muon spectrometer [41–44] and a forward hadron calorimeter [45], both located outside of the superconducting coil. One of the main goals of the CMS setup was to detect the Higgs boson by its decay mode into four leptons $H \rightarrow 4l$, where $l = e, \mu$, and the technical design of all the detector systems has led to the success in achieving this purpose.

The heart of the CMS detector–magnetic system comprises a NbTi superconducting solenoid and a magnetic flux return yoke made of construction steel containing up to 0.17% carbon and up to 1.22% manganese, as well as small amounts of silicon, chromium, and copper. The solenoid [34], creating a magnetic flux of 130 Wb, is installed in a vacuum cryostat in the central four-layer barrel wheel of the magnet yoke [35]. The magnet yoke also includes two three-layer barrel wheels around the solenoid cryostat on each side of the central barrel wheel, two nose disks inside each end of the solenoid cryostat, and four endcap disks on each side of the cryostat.

The inner diameter of the solenoid cryostat is 5.945 m, the diameter of each nose disk is 5.26 m, the inscribed outer diameters of the 12-sided barrel wheels around the cryostat are 13.99 m, the inscribed outer diameters of the 12-sided endcap disks are 13.91 m, the length of the magnet yoke is 21.61m, and the weight of the magnetic flux return yoke is more than 10,000 tons. There are two halves of the forward hadron calorimeter, and two radiation shields around the vacuum beam pipe downstream of the endcap disks on each side of the yoke at 10.86 m off the solenoid centre. The total thickness of steel along the radius of the central barrel wheel is 1.705 m, the total thickness of the endcap disks on each side is 1.541 m, and the thickness of each nose disk is 0.924 m.

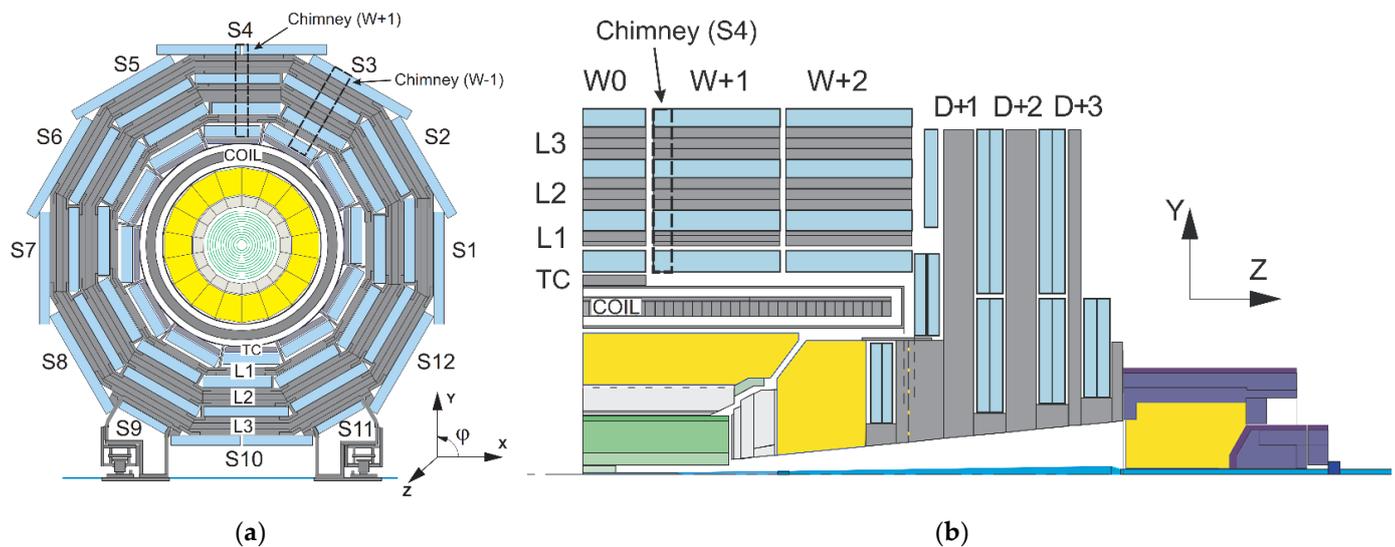

**Figure 1.** Schematic view [46] of the CMS detector in transverse (**a**) and longitudinal (**b**) sections. Designations are as follows: S1–S12 — the detector azimuthal sectors; TC and L1–L3 — layers of the yoke barrel wheels (W); D — the yoke endcap disks; Chimneys — grooves in the barrel wheels W+1 and W−1 for pipelines of the cryogenic system and current leads of the solenoid (COIL), respectively.

Only two thirds of the solenoid magnetic flux passes through the cross sections of the magnet yoke. The remaining third of the magnetic flux creates a stray magnetic field around the magnet yoke, the value of which decreases with a radius around the solenoid axis and with a distance along the axis. At a radius of 50 m from the coil axis in the central plane of the detector, the value of the magnetic flux density is 2.1 mT. At a distance of 50 m from the centre of the solenoid along its axis, the value of the magnetic flux density becomes 0.6 mT. The contribution of the barrel wheels, nose, and endcap disks to the central magnetic field is 7.97%, the contribution of the steel absorber of the forward hadron calorimeter, ferromagnetic elements of the radiation shielding, and the 40 mm thick steel floor of the experimental underground cavern is only 0.03%.

The CMS detector provides registration of charged particles in the pseudorapidity region $|\eta| < 2.5$, registration of *e* and $\gamma$ in the region $|\eta| < 3$, registration of hadronic jets in the region $|\eta| < 5.2$, and registration of $\mu$ in the region $|\eta| < 2.4$ [11]. The pseudorapidity $\eta$ is determined as $\eta = -ln[tan(\theta/2)]$, where $\theta$ is a polar angle in the detector reference frame.

The origin of the CMS coordinate system is located in the centre of the superconducting solenoid, the *X* axis lies in the LHC plane and is directed to the centre of the LHC machine, the *Y* axis is directed upward and is perpendicular to the LHC plane, and the *Z* axis makes up the right triplet with the *X* and *Y* axes and is directed along the vector of magnetic induction created on the axis of the superconducting coil.

Until 2013, a pixel tracking detector [37] consisted of 1440 silicon modules; a strip tracking detector [38] consists of 15,148 silicon modules. Both tracking detectors provide a resolution in the impact parameter of charged particles at a level of ≈15 μm and a resolution in the transverse momentum $p_T$ of charged particles of about 1.5% at $p_T$ = 100 GeV/*c*.

An electromagnetic homogeneous calorimeter ECAL [39] consists of 75,848 lead tungstate crystals and overlaps the pseudorapidity regions $|\eta| < 1.479$ in the central part (EB) and $1.479 < |\eta| < 3$ in the two endcaps (EE). In EB, the crystals have a length of 23 cm and an end surface area of 2.2 × 2.2 cm$^2$, while in EE, the length of the crystals is 22 cm, and the end surface area is 2.86 × 2.86 cm$^2$. All crystals are optically isolated from each other and are directed by their ends to the point of interaction of the primary particle beams, which allows the direction of $e$ и $\gamma$ to be determined and registered in the ECAL. A preshower strip detector is placed in front of each EE and has a layer of lead between two layers of sensitive elements equivalent to three radiation lengths. The preshower detector has a high spatial resolution that ensures the separation of two spatially close gamma rays.

The ECAL energy resolution for electrons with a transverse energy of about 45 GeV, originating from decays $Z \rightarrow e^+e^-$, is better than 2% in the central region at $|\eta| < 0.8$, and stands between 2% and 5% outside this area.

The hadron heterogeneous calorimeter (HCAL) [40] consists of brass layers of a hadron absorber and thin plates of plastic scintillators between them, which register signals from the passage of particles through the calorimeter. The register cells of the calorimeter are grouped along the calorimeter thickness in towers with projective geometry and granularity of $\Delta\eta \times \Delta\varphi = 0.087 \times 0.087$ in the central region (HB) at $|\eta| < 1.3$ and $\Delta\eta \times \Delta\varphi = 0.17 \times 0.17$ in the endcap area (HE) at $1.3 < |\eta| < 2.9$ (here $\varphi$ is an azimuthal angle in the detector reference system). The forward hadron calorimeter [45] increases the hadronic jet detection interval to the pseudorapidity of $|\eta| < 5.2$.

Finally, a muon spectrometer covering the pseudorapidity interval $|\eta| < 2.4$, consists of three systems for measuring muon momenta—drift tube chambers [41] in the central part of the muon spectrometer, cathode-strip chambers [42] in the endcap parts, and resistive plate chambers [43]. The global fit of the muon trajectory in the muon spectrometer to the parameters of the muon trajectory found and reconstructed in the tracking detectors provides a resolution in the particle transverse momentum, averaged over the azimuthal angle and pseudorapidity, at a level from 1.8% for the muon transverse momentum of $p_T$ = 30 GeV/$c$ to 2.3% for $p_T$ = 50 GeV/$c$ [44].

*2.2. Formulation of the Magnetostatic Problem for Modelling the Magnetic Flux of the CMS Superconducting Solenoid*

A three-dimensional model of the CMS magnetic system reproduces the magnetic flux generated by the system in a cylindrical volume of 100 m in diameter and 120 m in length [47–53]. The well-proven TOSCA (two scalar potential method) program [54], developed in 1979 [55] in the Rutherford Appleton Laboratory, is chosen as a tool for creating a model of the magnetic system.

It is well known [56] that the problem of determining the magnetic field of linear currents, neglecting the volumes of conductors, can be solved as a problem of potential theory. The main idea of the TOSCA program is to use nonlinear partial differential equations with two scalar magnetic potentials for solving magnetostatic problems—total (in the Laplace equation) and reduced (in the Poisson equation) [57]. For this, two regions are distinguished in the magnetic system model. In one, $\Omega_j$, containing conductors with a direct current, the reduced scalar magnetic potential $\varphi$ is used for the solution, as well as the Biot–Savard law [56] which takes into account the magnetic field of current-carrying conductors; in the other, $\Omega_k$, which does not contain current-carrying conductors, but contains ferromagnetic isotropic or anisotropic materials, the total scalar magnetic potential $\psi$ is used for the solution, and at the interface between the two regions the normal components of the magnetic flux density $B_n$ and the tangential components of the magnetic field strength $H_t$ satisfy the continuity conditions [56,58] as follows:

$$B_{nk} = B_{nj}, \tag{1}$$

$$H_{tk} = H_{tj}. \tag{2}$$

Herein, at the remote outer boundaries of the domain $\Omega_k$ depending on the configuration of the magnetic system, either the Dirichlet $\psi = 0$ or Neumann $\frac{\partial \psi}{\partial n} = 0$ boundary conditions are used, where $n$ is the outer unit normal to the boundary of the domain $\Omega_k$.

The basic equations for solving the nonlinear magnetostatic problem are Maxwell's equations [56]

$$\nabla \cdot \boldsymbol{B} = 0, \tag{3}$$

$$\nabla \times \boldsymbol{H} = \boldsymbol{J}, \tag{4}$$

where vector $\boldsymbol{B}$ denotes the magnetic flux density, vector $\boldsymbol{H}$ denotes the magnetic field strength, and $\boldsymbol{J}$ is the vector of a given current density in the current elements of the magnetic system. Herein, the vectors $\boldsymbol{B}$ and $\boldsymbol{H}$ are related to each other by the equation

$$\boldsymbol{B} = \mu(\boldsymbol{H})(\boldsymbol{H} - \boldsymbol{H}_c), \tag{5}$$

where $\mu(\boldsymbol{H})$ is the magnetic permeability of the medium in which the magnetic field is determined, and $\boldsymbol{H}_c$ is the coercive force in the medium.

For a nonlinear problem $\mu(\boldsymbol{H})$ is a function of the magnetic field strength in the medium and in general case can be a tensor. For isotropic materials such as construction steel used in the CMS magnet yoke,

$$\mu(|\boldsymbol{H}|) = \mu_r(|\boldsymbol{H}|)\mu_0, \tag{6}$$

where $\mu_0$ is the magnetic permeability of free space, and $\mu_r(|\boldsymbol{H}|)$ is the relative magnetic permeability, which in a magnetic medium is a dimensionless nonlinear function of the magnetic field strength and reaches a value of about 2000 units in steel. Equation (5) is the so-called material magnetization curve. At the same time, the coercive force $\boldsymbol{H}_c$ in the most magnetic materials is assumed to be zero, but it plays an essential role in permanent magnets.

By the Helmholtz expansion theorem, if the divergence and curl of a vector field are defined at each point of a finite open region in space, then everywhere in this region the vector field can be represented as a sum of irrotational and solenoidal fields. Then, in the region $\Omega_k$, where there are no current elements, the vortex part of the magnetic field strength is absent and the field $\boldsymbol{H}$ is solenoidal and can be represented as the gradient of the total scalar potential $\psi$ at any point in this region:

$$\boldsymbol{H} = -\nabla \psi. \tag{7}$$

Thus, considering Equations (5), (6), Equation (3) transforms into the Laplace equation for the scalar potential $\psi$:

$$\nabla \cdot \mu_r \nabla \psi = 0. \tag{8}$$

In the domain $\Omega_j$, containing conductors with a direct current, the vector of the magnetic field strength is separated into two parts—the solenoidal field $\boldsymbol{H}_m$ represented as the gradient of the reduced scalar potential $\varphi$

$$\boldsymbol{H}_m = -\nabla \varphi, \tag{9}$$

and the vortex field of the current elements $\boldsymbol{H}_s$, which, at a distance $\boldsymbol{R}$ from the current source, is determined according to the Biot–Savard law:

$$\boldsymbol{H}_s = \int_{\Omega_j} \frac{\boldsymbol{J} \times \boldsymbol{R}}{|\boldsymbol{R}|^3} d\Omega_j. \tag{10}$$

Thus, considering Equations (5), (6), (9), and (10), Equation (3) transforms into the Poisson equation for the reduced scalar magnetic potential $\varphi$:

$$\nabla \cdot \mu_r \nabla \varphi = \nabla \cdot \mu_r \boldsymbol{H}_s. \tag{11}$$

Using the full scalar magnetic potential in the space around conductors significantly increases the accuracy of calculations. Both Equations (8) and (11) are solved in the TOSCA program by the numerical finite element method [59] at nodes of a mesh, which discretises the entire region of the magnetic system model into quadrangular and triangular prisms. The introduction of two types of scalar magnetic potentials in the TOSCA program is determined by the fact that the fields $\boldsymbol{H}_m$ and $\boldsymbol{H}_s$ in the ferromagnetic elements of the magnetic system outside the conductors are directed towards each other, and the calculation of the resulting magnetic field strength by the method of the reduced scalar potential leads to large errors. The use of the total scalar magnetic potential in the space around conductors significantly increases the accuracy of calculations.

Thus, to solve the magnetostatic problem in the TOSCA program [54] means to calculate the scalar magnetic potential, total or reduced, at the nodes of the spatial finite element mesh. The components of the magnetic field strength $\boldsymbol{H}$ are then calculated as the gradients of the scalar potentials (Equations (7) and (9)), while in the region of the reduced scalar potential they are supplemented with the components of the magnetic field strength created by current-carrying conductors and determined by the Biot–Savard law (10). The components of the magnetic flux density $\boldsymbol{B}$ are calculated using Equations (5) and (6) under the conditions of Equations (1) and (2) of the component continuities.

## 3. Results

### 3.1. Description of the CMS Superconducting Solenoid Model

In the model of the CMS magnetic system, the region of the reduced scalar potential $\Omega_j$ is a system of five cylinders with a total length along the Z axis of 12.666 m and with diameters from 6.94625 to 6.95625 m. The central cylinder has the largest diameter, the diameter of the two adjacent cylinders is 4 mm less, and the diameter of the two outer cylinders is another 6 mm less than the diameter of the central cylinder. This configuration of the volume containing the superconducting coil reflects the deformation of the solenoid under electromagnetic forces at an operating current of 18.164 kA, which leads to an increase in the coil radius by 5 mm in the central plane of the solenoid [33]. The axial length of the volume, equal to 12.666 m, corresponds to the distance between the nose discs of the magnet yoke at the operating current of the solenoid. The rest of the CMS magnet model volume is the region $\Omega_k$ of the total scalar potential $\psi$ which is also distributed inside the ferromagnetic elements of the system. The entire CMS magnet model is subdivided for linear finite elements with azimuth lengths corresponding to an angle of 3.75°. In the region of the reduced scalar potential $\varphi$, the average length of the finite element is 65.5 mm in the radial direction, and 86.8 mm in the axial direction.

The CMS superconducting solenoid winding consists of four layers of a superconductor with a cross section of 64 × 21.6 mm$^2$, stabilized with pure aluminium and wound by the short side on the inner surfaces of five modules of a mandrel made of 50 mm thick aluminium alloy. The inner diameter of the mandrel is 6.846 m, the axial length of each module is 2.5 m, and the radial thickness of the solenoid, including the mandrel, interturn, and interlayer electrical insulation, is 0.313 m [34]. In the direction of the magnetic field inside the solenoid along the Z axis, the modules are labelled as CB−2, CB−1, CB0 (central module), CB+1, and CB+2. Each superconductor layer in one coil module consists of 109 turns, except for the inner layer of the CB−2 module, in which only 108 turns are wound. The total number of turns in the solenoid is 2179, the total current strength in the coil at an operating current of 18.164 kA is 39.58 MA-turns. The loss of one turn during winding the CB−2 module led to a shift in the geometrical position of the magnetic flux density maximum central value with respect to the origin of the CMS coordinate system by 16 mm in the positive direction of the Z axis, which was then taken into account in the operation of tracking detectors.

The superconductor is composed of a superconducting cable with a cross section of 20.63 × 2.34 mm², an aluminium stabilizer around it with a purity of 99,998% and a cross section of 30 × 21.6 mm², and two reinforcements of a high-strength aluminium alloy with cross section of 17 × 21.6 mm² [60] each, welded to the sides of the aluminium stabilizer by electron beam welding. The superconducting cable is composed of 32 superconducting twisted strands of 1.28 mm in diameter, each of which contains from 500 to 700 filaments of superconducting NbTi alloy extruded in a high purity copper matrix [61]. All dimensions are given at room temperature of 295 K. In the solenoid computing model, the position of the superconducting cable corresponds to the temperature contraction of the linear dimensions with a factor of 0.99585 when the solenoid is cooled down with liquid helium to an operating temperature of 4.2 K.

When describing the solenoid in the CMS magnet model, in correspondence with an assumption that at the achieved superconductivity all the current flows only through the superconducting cable, only the cable geometrical dimensions and location at cryogenic temperature are considered. Each coil module, with the exception of CB−2, is presented in the form of four concentric cylinders of 2.4532 m long and 20.54 mm thick with average diameters corresponding to the deformation of the solenoid when the operating current reaches a value of 18.164 kA [48–50]. So, for example, in the central module CB0, the average diameters of the current cylinders are 6.37008, 6.50034, 6.6306, and 6.76086 m. The average diameters of the CB±1 current cylinders are each 4 mm less and in modules CB±2 the current cylinder bores are reduced by another 6 mm. The distance between the cylinders of adjacent modules in the model is 41.4 mm. The inner layer of the CB−2 module in the model consists of two cylinders with an average diameter of 6.36008 m—a single turn of 2.33 mm long nearby the CB−1 module, and a cylinder of 2.4158 m long separated from this turn by an air gap of 35.07 mm. The current density in the cylinders is calculated from the number of Ampere-turns in the section of each cylinder. Its value has three meanings—one is for each of 19 cylinders with the same cross-sectional area and the two others are for cylinders corresponding to a single and 107 turns of a superconducting cable, respectively. In addition to the current cylinders, the model also includes two current linear conductors corresponding to the coil electrical leads. These conductors are also surrounded in the model by a region of reduced scalar magnetic potential.

An accurate description of the geometry of the coil conductors in the model plays an essential role in the reconstruction of the trajectories of muons penetrating the entire thickness of the coil winding. So, for example, in the middle plane of the CB0 module inside the solenoid winding, the magnetic flux density decreases stepwise from 2.94 T between the first and second layers to 0.93 T between the third and fourth layers, which should be taken into account when reconstructing the muon trajectory.

*3.2. Description of the CMS Magnet Yoke Model*

In Figure 2 an isometric view of the CMS magnetic flux return yoke model is shown. As ferromagnetic elements of the magnet yoke, the model includes five multi-layer steel barrel wheels, W0, W±1, and W±2, around the solenoid cryostat, steel nose disks, four endcap disks, D±1, D±2, D±3, and D±4, at each side of the solenoid cryostat, steel brackets for connecting the layers of the barrel wheels, steel feet of the barrel wheels and carts of the endcap disks, steel absorbers and collars of the forward hadron calorimeter, steel elements of the radiation shielding, and collimators of proton beams, as well as the 40 mm thick steel floor of the experimental underground cavern with an area of 48 × 9.9 m² [48–53]. Herein, the signs "+" and "−" denote the barrel wheels and endcap disks at positive and negative values of coordinates along the Z axis, respectively.

According to the number of faces, the barrel wheels and endcap disks have 12 azimuthal sectors, 30° each. The sectors are numbered in the direction of increasing the values of the azimuthal angle and the numbering starts from a horizontally located sector S1, the middle of which coincides with the *X* axis. As can be seen from Figure 3a, each barrel wheel sector consists of three layers, connected by steel brackets, one (L1) of 0.285 m thick and two (L2 and L3) of 0.62 m thick each. Thick layers consist of two types of steel—steel *G* is in 0.085 m thick linings and steel *I* is in a core with a thickness of 0.45 m.

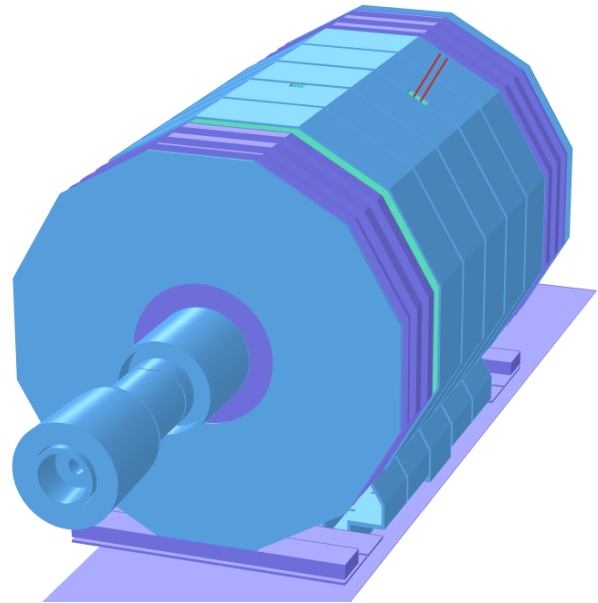

**Figure 2.** Three-dimensional model of the CMS magnet [53], using the TOSCA program for calculating the magnetic flux at a solenoid current of 18.164 kA. The ferromagnetic materials of the magnet yoke are marked with different color shades, in which three different magnetization curves are used. Shown are five steel barrel wheels, W0, W±1, and W±2, with their feet; four steel endcap disks, D±1, D±2, D±3, and D±4, on each side of the central part with the upper parts of their carts; the most distant ferromagnetic elements of the model, extending to distances of ±21.89 m in both directions from the centre of the solenoid; steel absorbers and collars of the forward hadron calorimeter; ferromagnetic elements of the radiation shielding; and collimators of the proton beams. Two linear conductors of the current leads in a special chimney in the W−1 barrel wheel, a chimney in the W+1 barrel wheel for pipelines of the cryogenic system, and a 40 mm thick steel floor are seen as well.

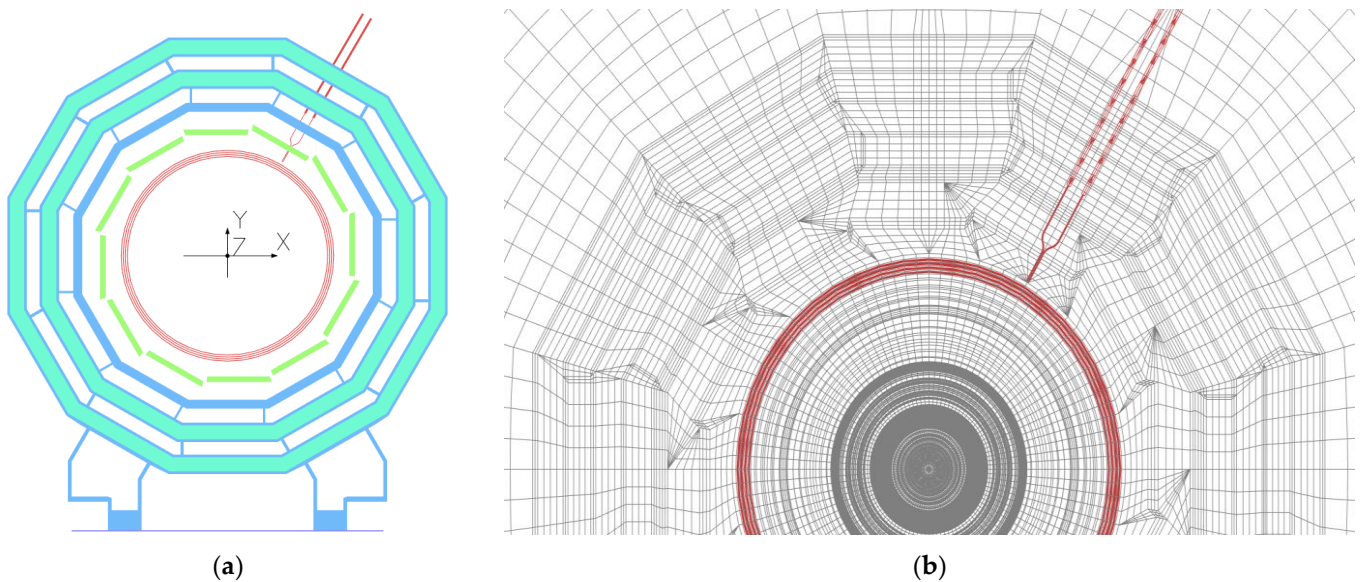

(**a**)      (**b**)

**Figure 3.** (**a**) Model of the central barrel wheel W0 with steel feet and a floor in the presence of four layers of superconducting cable and two linear current conductors and (**b**) finite element mesh in the *XY* plane.

The distance between the thin layer L1 and the middle thick layer L2 is 0.45 m, the same between two thick layers is 0.405 m. The air gaps between the central barrel wheel W0 and the adjacent wheels W±1 are 0.155 m each, the gaps between the wheels W±1 and W±2 are equal to 0.125 m each [50]. There is an additional fourth TC layer of 0.18 m thick in the central wheel W0 at 3.868 m from the solenoid axis, which absorbs hadrons escaping the barrel hadron calorimeter due to its insufficient thickness at low values of pseudorapidity. The blocks of this layer are displaced in the azimuth angle by 5° with respect to other blocks in the sector [51]. The distance between the layers TC and L1 is 0.567 m, and their material is steel *G*. The same type of steel is used in the connecting brackets and the barrel wheel feet. The material for the floor of the experimental underground cavern is steel *S*.

Steel *S* is used in the large and small endcap disks that close the flux-return yoke on each side of the solenoid cryostat, as well as in the connection rings between them and in the plates of the disk carts. The thickness of each of the first two disks is 0.592 m, the third disk is 0.232 m thick, and the fourth small disk is 0.075 m. The air gaps between the W±2 barrel wheels and the D±1 disks are 0.649 m each, the gaps between the D±1 and D±2 disks are 0.663 m each, the gaps between the D±2 and D±3 disks are 0.668 m each and, finally, the gaps between the D±3 and D±4 disks are 0.664 m each [50].

In 2013–2014 the fourth small disk of 5 m in diameter was increased to an inscribed diameter of 13.91 m, and the thickness of this outer part of the disk is 0.125 m [51]. The extended part of the fourth disk consists of two steel plates, each 25 mm thick, and a specialized concrete between them containing oxides of boron and iron, while the iron content is 57%. In the model the magnetization curve of steel *G* is used for the fourth disk's extended plates, and concrete is described by the same curve with a packing factor of 0.57.

When building a three-dimensional model, a two-dimensional mesh of finite elements is created initially in the *XY* plane, as shown in Figure 3b, where all ferromagnetic elements used in the model are discretised. Then this mesh is extruded layer-by-layer in the direction of the *Z* axis, while the coordinates of the mesh nodes are transformed to describe complex geometric volumes, mainly cylindrical and conical, with a minimum number of the plane mesh nodes. Layers in the *Z* direction and the location of elements in the *XY* planes between them are used to describe the materials of the model elements. At present, the model of the CMS magnetic system contains 140 layers and 8,759,730 nodes of the spatial mesh in a cylindrical volume with a diameter of 100 m and a length of 120 m.

### 3.3. Magnetization Curves of Steel in the CMS Magnet Model

To describe the properties of the ferromagnetic elements of the CMS magnetic system, the model uses three curves of the isotropic nonlinear dependence of the magnetic flux density ***B*** on the magnetic field strength ***H*** [49], presented in Figure 4 on a semi-log scale.

Each *B-H* curve was obtained by averaging the magnetization curves measured for samples corresponding to different melts of a given type of steel used in the flux-return yoke elements. For the magnetization curve of steel *G*, averaging was performed over 33 samples, while scattering between the magnetization curves of the samples was, on average, $(11.5 \pm 9.1)\%$. For the magnetization curve of steel *I*, averaging was performed over 65 samples, and a spread between the magnetization curves of the samples was, on average, $(8.7 \pm 8.0)\%$. For the magnetization curve of steel *S* averaging was performed over 72 samples, while a dispersion between the magnetization curves of the samples was, on average, $(8.2 \pm 7.8)\%$.

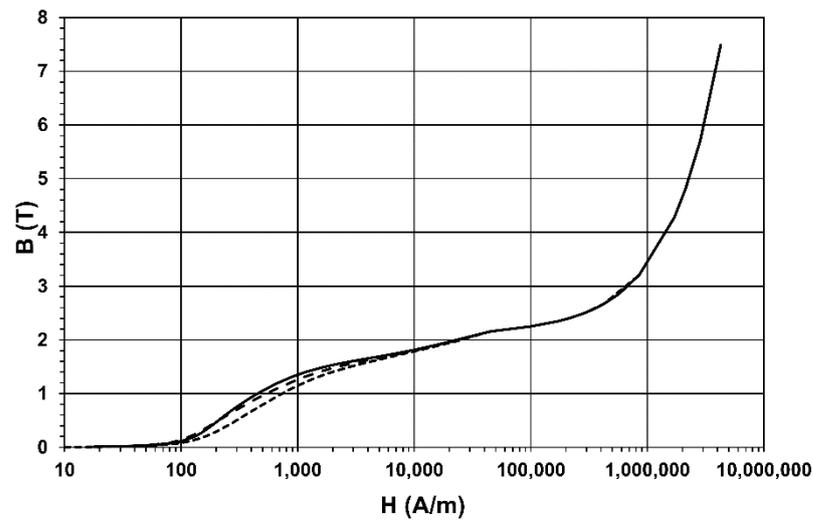

**Figure 4.** Magnetization curves of the steel used in the CMS magnet. Steel *I* (dashed line) is used to produce blocks of the barrel wheel thick layers. Steel *S* (dotted line) is used to produce the yoke nose and endcap disks. The same *B–H* curve in the model describes the magnetic properties of the disk carts and the floor of the experimental cavern. The magnetization curve of steel *G* (solid line) is used in the model for all other ferromagnetic elements.

The magnetization curves of steel samples were measured in the range *B* from 0.003 to 2 T. For the magnetic flux density values exceeding 2.15 T, all curves use the magnetization curve of steel *I*, measured up to a *B* value of 7.4887 T. In this interval, the dependence of *B* vs. *H* becomes linear with a slope coefficient greater than the magnetic permeability of vacuum by 0.42%.

## 4. Discussion

The development of the CMS magnet model strongly depended from the computer technology capabilities. The first model, shown in Figure 5, was created in June 1997 in the framework of the TOSCA program version 6.6 and was calculated on a UNIX server IBM RISC System 6000. The scalar magnetic potential was calculated in 1/8 part of a cylindrical volume with a radius of 13 m. The length of the part was 14 m. The model included 20 cylinders of a superconducting cable with dimensions corresponding to a temperature of 4.2 K and increased by 5 mm to consider the electromagnetic forces acting on the solenoid. The model comprised a half of the barrel wheels and four endcap disks described by one *B-H* curve, since the production of the yoke blocks had not yet begun. The model contained 83,832 spatial mesh nodes and was calculated for 1.9 h of CPU time. Today, this model required only 56 s of processor time to be calculated on a personal computer with a dual-core processor having a clock rate of 2.4 GHz.

A limitation on the maximum number of nodes to 400,000 in the TOSCA program version 6.6 did not allow the volume of the next model to be expanded by more than to half a cylinder with a radius of 13 m and a length of 28 m, but this extension was sufficient to calculate the mechanical forces on a superconducting solenoid at its possible displacements under the acting electromagnetic forces [62,63], that allowing the scheme of the coil suspension in the cryostat to be selected.

A year later, the first basic CMS magnet model, version 66_980612 [47,64], which included the steel elements of the forward hadron calorimeter, was created. This model allowed the influence of various configurations of the CMS magnet yoke on the field inside the solenoid to be investigated. In particular, from the magnetic field map shown in Figure 6 and from the calculation of axial forces acting on the yoke elements, the necessity to manufacture a support cylinder for the endcap calorimeters from non-magnetic stainless steel was declared. The finite-element mesh was constructed from 107,784 nodes.

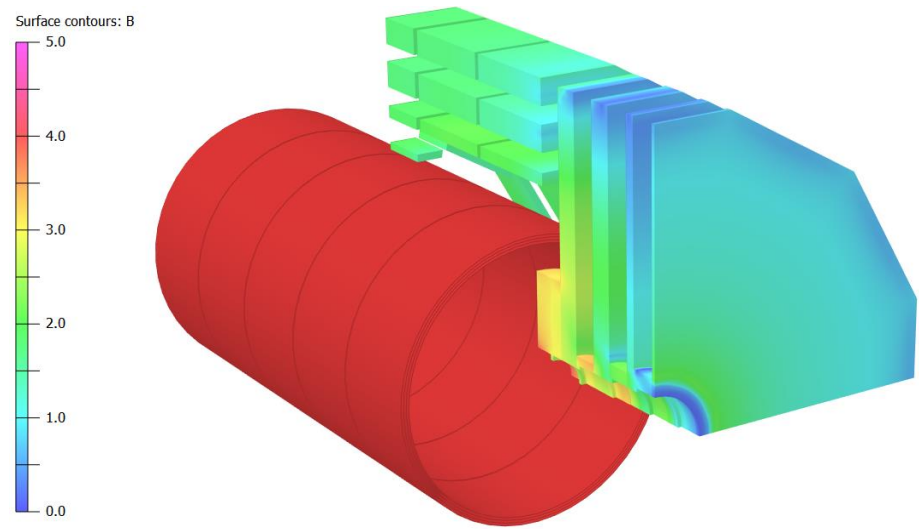

**Figure 5.** The first model of the CMS magnet, developed in 1997 and containing 83,832 nodes of the spatial finite element mesh.

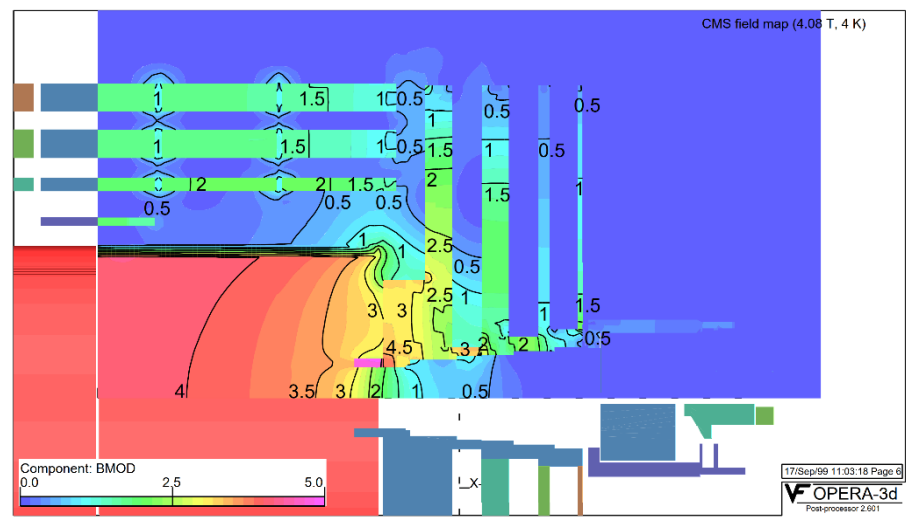

**Figure 6.** The CMS magnetic field map calculated with a solenoid current of 19.5 kA [47]. The numbers indicate the values of the magnetic flux density *B* in T.

Version 66_001012 became a basic model for the next three years. This model allowed the influence of layers L2 and L3 thin plates' recession (by 36 mm with respect to the thick inner blocks) on the distribution of magnetic flux in the gaps between the barrel wheels, to be studied. In this model, the scalar magnetic potential was calculated in 1/24 part of a cylindrical volume with a radius of 13 m. The length of the part was 18 m. To obtain the distribution of the magnetic flux in a full cylinder volume of 26 m in diameter and 36 m in length, 12 rotations of the model 30° azimuthal sector around the *Z* axis, and a reflection of the scalar potential values with a minus sign into the region of negative coordinates of the *Z* axis with respect to the *XY* median plane were used. In contrast to the previous version, three *B–H* curves described the types of steel that were used in the flux-return yoke. The model was calculated on Sun-4u computers with the 64-bit UltraSPARC processors on a mesh of 219,678 spatial nodes

The next model version, 85_030919, differed from the previous version, 66_001012, only by the number of mesh nodes, which increased in the TOSCA program version 8.5 to 822,492. This model, for the first time, allowed to create a space map of the magnetic flux distribution in the entire volume of the CMS detector. The idea of this map came from earlier works [64,65] and was based on discretizing the entire space of the magnetic system into primitive volumes of several types, such as sectors of cylinders and cones, tubes and truncated tubes, and prisms and parallelepipeds. In the CMS magnet model, all rotation volumes are centred along the Z axis; the barrel wheel elements are described by prisms and parallelepipeds. This arrangement of the volumes allows each azimuthal sector of the magnet model to cut the rotation volumes by YZ planes with a constant step along the azimuthal angle, and to cut the prisms and parallelepipeds by planes parallel to the outer faces of the barrel wheels with a constant step between them. Each cutting plane can be evenly divided into cells, and, thus, each primitive volume contains a mesh of the space nodes.

In these nodes three components of the magnetic flux density are calculated from the scalar potential distribution obtained with TOSCA program, and the field between the nodes is computed by a linear interpolation of the magnetic induction values over the eight nearest nodes [64–66]. The discretization of space into primitive volumes allows a fast search for the needed nodes around the coordinates of the points along the trajectories of charged particles to be performed.

Along the trajectories of charged particles in the programs for simulation and reconstruction of the primary particle collision events, the special class *MagneticField* returns the values of the magnetic flux density components for the coordinates of each requested space point. To describe the magnetic field map calculated with the model version 85_030919, 271 primitive volumes in the only azimuth sector S1 were used. The magnetic field in all other sectors was determined by rotating the coordinates of the desired point to sector S1 and then by translating the components of the found magnetic induction to the desired space point.

At the beginning of 2007, the TOSCA program version 11.03 appeared and then in May 2007, it was installed at CERN on the 64-bit x86_64 processor computers under the Linux operating system. In this version, the limit on the mesh node number was increased to 5 million, which allowed the magnetic flux with model version 1103_071212 in a half cylinder volume with a radius of 13 m and a length of 40 m to be described. The model, shown in Figures 7 and 8, contained 1,922,958 nodes of the spatial mesh and included the deformation of the solenoid and a displacement of the nose and endcap disks toward the coil centre under the electromagnetic forces as well as an absence of one turn in the inner layer of the CB–2 coil module.

The model comprised the barrel wheel steel feet, the upper plates of the endcap disk carts, and an inner part of the radiation shielding. The new CMS magnetic field map used 312 primitive volumes to describe the magnetic flux distribution in the azimuthal sector S1. Distinguishing it from the previous version of the magnetic field map, the new map included the sectors S3 and S4 with chimneys for cryogenic and electrical leads (see Figure 1), as well as the sector S11 with the barrel wheel feet. The magnetic induction in the remaining sectors but S9 was determined based on the primitive volumes of the sector S1.

The simulated magnetic field values were compared with the field values obtained in 2006 [36] with a field mapping machine [67] measuring the magnetic flux density $B$ inside the superconducting solenoid. The measurements were performed with an accuracy of 0.07%. The model reproduced the measured field with an accuracy of 0.1%. In this model, for the first time, the coil operating current of 18.164 kA was used, which gave a central magnetic flux density of 3.81 T.

At the end of 2008, the CMS detector was assembled in an underground experimental cavern and commissioned with the cosmic muons. It was noted [46], that the magnitude of the magnetic field in the layers of the barrel wheels in the model, especially the extreme ones, is overestimated by several percent. This effect of the magnetic field map arose because of a compression of the returned magnetic flux of the solenoid in the insufficient full volume of the model.

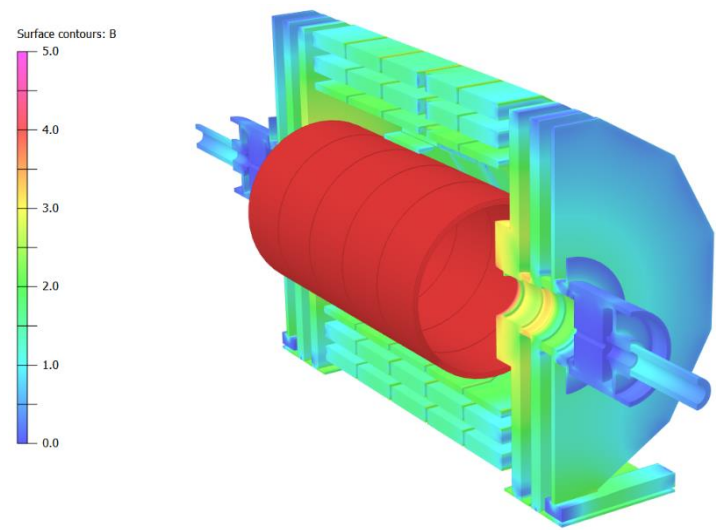

**Figure 7.** The CMS magnet model version 1103_071212. The colour scale corresponds to the interval of the magnetic flux density *B* from zero to 5 T with an increment of 0.5 T. The distribution of the magnetic field is presented on the inner surfaces of the flux-return yoke.

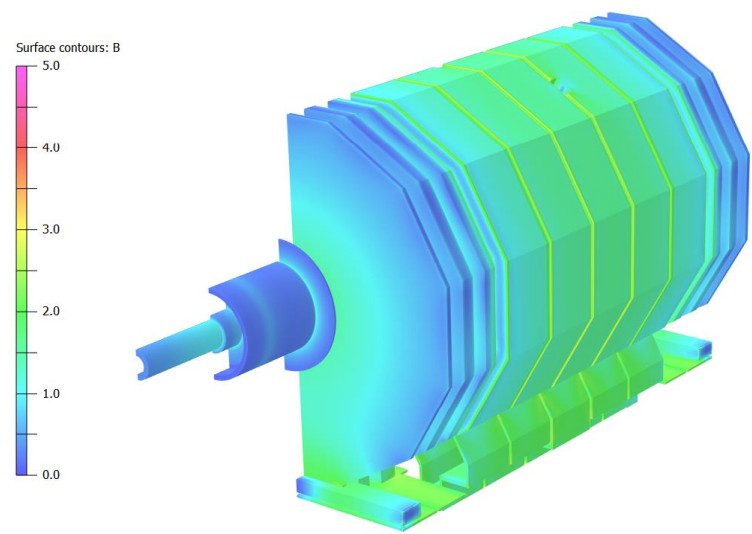

**Figure 8.** The CMS magnet model version 1103_071212. The colour scale corresponds to the interval of the magnetic flux density *B* from zero to 5 T with an increment of 0.5 T. The distribution of the magnetic field is presented on the outer surfaces of the flux-return yoke.

To minimize this effect, a half a cylinder used for calculations was expanded by increasing the cylinder radius from 13 to 30 m and the cylinder length from 40 to 70 m. Consequently, the number of spatial mesh nodes in the new model 1103_090322 was increased to 1,993,452. The model comprised a 40 mm thick steel floor of the underground experimental cavern. The magnetic field map created with this model was used in the CMS detector during the entire first run of the LHC operation in 2009–2012, and with this description of the magnetic flux distribution in the detector volume the Higgs boson was discovered [2,3].

In the next version, 14_120812, both halves of the CMS magnet yoke were separately included in the model. The model comprised enlarged fourth endcap disks, the endcap disk keels in the azimuthal sector S10, and an outer part of the radiation shielding with cylindrical gaps between the shield and the collar of the forward hadron calorimeter. The volume of each half a cylinder used for calculations was increased in radius from 30 to 50 m and in length from 70 to 120 m. This version was the last one designed for the computers based on the x86_64 processor, and the division of the model into two halves was dictated by the limitation on the number of mesh nodes. One of the model halves contained 3,624,593 nodes, the other one used 3,584,357 nodes.

To create the magnetic field map, 720 primitive volumes were used in each azimuthal sector divided into two halves of 15° azimuth angle each. The total number of primitive volumes in a cylinder with a diameter of 18 m and a length of 40 m, used to describe the distribution of the magnetic flux in the CMS detector, was 8640.

Both halves of the model 14_120812 were combined into a single volume in the next model version, 16_130503, shown in Figure 9. Starting with this version, all subsequent models were designed for personal computers with an Intel64/x64 processor and with RAM of 8–16 GB under the Windows operating system. The model contained 7,111,713 nodes of the spatial mesh and required 10 h of processor time to calculate. To create the magnetic field map, the number of primitive volumes in a cylinder with the same dimensions of 18 m in diameter and 40 m in length was increased to 9648 due to a more detailed description of the regions outside the barrel wheels, where the last layer of the muon drift tube chambers is located.

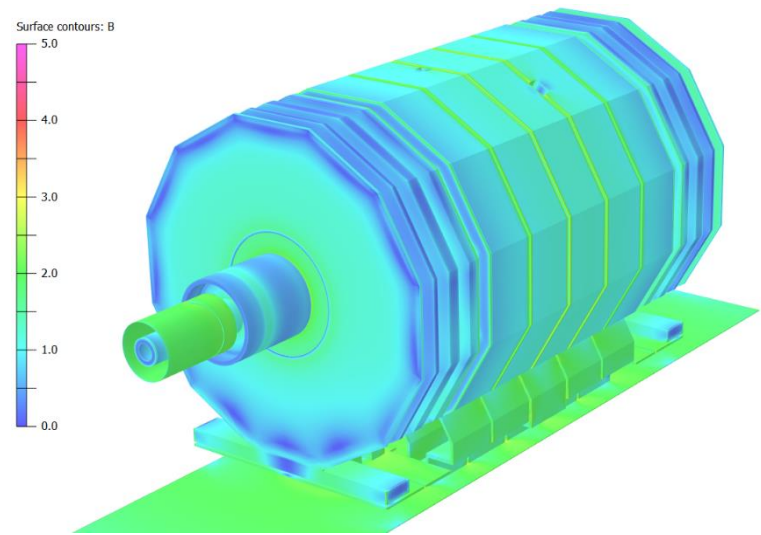

**Figure 9.** The CMS magnet model version 14_130503. The colour scale corresponds to the range of magnetic flux density *B* from zero to 5 T with an increment of 0.5 T. The distribution of the magnetic field is shown on the outer surface of the flux-return yoke.

Finally, the next two versions, 18_160812 and 18_170812 of the model differ from each other only by a slight refinement of the magnetization curve of steel *S*, which, like the other curves, did not change after version 66_001012. To complete the geometrical description of the magnetic system, the most distant part of the radiation shielding was added to both model versions propagated to distances of ±21.89 m on both sides from the centre of the solenoid, as shown in Figure 2. Thus, the length of the CMS magnetic field map cylinder was increased to 48 m, and the number of primitive volumes in all 12 azimuthal sectors was increased to 11,088.

## 5. Conclusions

For more than 10 years of the CMS detector operation, a computing modelling of the magnetic system has been used to obtain the distribution of the magnetic flux throughout the experimental setup. Applying the magnetic flux simulation to the CMS magnet, results in the creation of a magnetic field map in the entire CMS volume, allowing the precise measurement of the electron (positron) and muon momenta, which has made it possible to reconstruct the invariant mass of the Higgs boson—the last brick of the Standard Model of elementary particles—with a high precision.

Several consecutive versions of the CMS magnet model were developed using different types of geometrical and scalar magnetic potential symmetries. This has allowed the magnetic flux in the entire CMS detector during a long period of time to be described with minimal computer resources, corresponding to the available computer technology capabilities.

**Funding:** This research received no external funding.

**Institutional Review Board Statement:** Not applicable.

**Informed Consent Statement:** Not applicable.

**Data Availability Statement:** Not applicable.

**Acknowledgments:** The author is very thankful to Alain Hervé, François Kircher, Richard P. Smith, Austin Ball, Wolfram Zeuner, Eduard Boos, Lyudmila Sarycheva, and Mikhail Panasyuk for administrative and technical support for many years.

**Conflicts of Interest:** The author declares no conflict of interest.